\documentclass[twocolumn]{aastex631}

\usepackage{xspace}

\newcommand{\CaII}{Ca\,\footnotesize II\normalsize\xspace}
\newcommand{\Mgb}{Mg\ $b$\xspace}
\newcommand{\kms}{\ \rm km \ s^{-1}}
\newcommand{\Msun}{\ \rm M_{\scriptscriptstyle\odot}}
\newcommand{\MsunLsun}{\ \rm M_{\scriptscriptstyle\odot} \ L_{\scriptscriptstyle\odot,H}^{-1}}

\begin{document}

\title{A Stellar Dynamical Mass Measurement of the Supermassive Black Hole in NGC 3258}

\correspondingauthor{Thomas K. Waters}
\email{waterstk@umich.edu}

\author[0000-0002-5231-7240]{Thomas K. Waters}
\affiliation{University of Michigan $\vert$ Department of Astronomy\\
1085 S University Ave, \\
Ann Arbor, MI 48109, USA}

\author[0000-0002-1146-0198]{Kayhan Gültekin}
\affiliation{University of Michigan $\vert$ Department of Astronomy\\
1085 S University Ave, \\
Ann Arbor, MI 48109, USA}

\author[0000-0002-8433-8185]{Karl Gebhardt}
\affiliation{University of Texas at Austin $\vert$ Department of Astronomy\\ 
2515 Speedway, \\
Austin, TX 78712, USA}

\author[0000-0001-6920-662X]{Neil Nagar}
\affiliation{Universidad de Concepción $\vert$ 
Departamento de Astronomía \\
Avenida Esteban Iturra s/n \\
Casilla 160-C \\
Concepción, Chile}

\author[0009-0002-0460-0436]{Vanessa Ávila}
\affiliation{Universidad de Concepción $\vert$ 
Departamento de Astronomía \\
Avenida Esteban Iturra s/n \\
Casilla 160-C \\
Concepción, Chile}
\shorttitle{NGC 3258 Stellar Dynamical SMBH Mass}
\shortauthors{Waters et al.}

\received{April 4, 2024}
\revised{June 7, 2024}
\accepted{June 19, 2024}

\begin{abstract}

    We present a stellar dynamical mass measurement of the supermassive black hole in the elliptical (E1) galaxy NGC 3258. Our findings are based on integral field unit spectroscopy from the Multi Unit Spectroscopic Explorer (MUSE) observations in narrow-field mode with adaptive optics and the MUSE wide-field mode, from which we extract kinematic information by fitting the \CaII and \Mgb triplets, respectively. Using axisymmetric, three-integral Schwarzschild orbit library models, we fit the observed line-of-sight velocity distributions to infer the supermassive black hole mass, the $H$-band mass-to-light ratio, the asymptotic circular velocity, and the dark matter halo scale radius of the galaxy. We report a black hole mass of $(2.2 \pm 0.2)\times10^9 \Msun$ at an assumed distance of $31.9 \ \rm Mpc$. This value is in close agreement with a previous measurement from Atacama Large Millimeter/submillimeter Array CO observations. The consistency between these two measurements provides strong support for both the gas dynamical and stellar dynamical methods.

\end{abstract}


\section{Introduction} \label{sec:intro}

    Supermassive black holes (SMBHs), which have masses greater than approximately $10^6 \ \rm M_\odot$, are integral parts of potentially all massive galaxies \citep{Kormendy&Ho2013, Kormendy&Gebhardt2001, Magorrian2018, Richstone1998}. Although SMBHs have very small spheres of influence (less than 1 $\rm kpc$), defined as the region within which the enclosed stellar mass is equal to the mass of the central SMBH, they exert profound impacts on the properties of their host galaxies \citep{Ricci2017_soi}. SMBH masses, in particular, show strong correlations with stellar velocity dispersions  \citep[$\sigma$;][]{Ferrarese&Merritt2000_msigma, Gebhardt2000_msigma}, total stellar mass \citep[$M_*$;][]{Reines2015_stellarmass}, bulge mass  \citep[$M_\mathrm{bulge}$;][]{Kormendy&Richstone1995_luminosity}, luminosity  \citep[$L$;][]{Dressler1989_luminosity, Kormendy1993_luminostity, Kormendy&Richstone1995_luminosity, Magorrian2018}, dark matter halo mass \citep[$M_{\mathrm{DM}}$;][but see also \citealp{2011Natur.469..374K}]{2002ApJ...578...90F, 2011ApJ...737...50V, 2024ApJ...960...28V} and X-ray luminosity  \citep[$L_X$;][]{Gaspari2019_xraylum}.

\begin{figure*}[ht]
\plotone{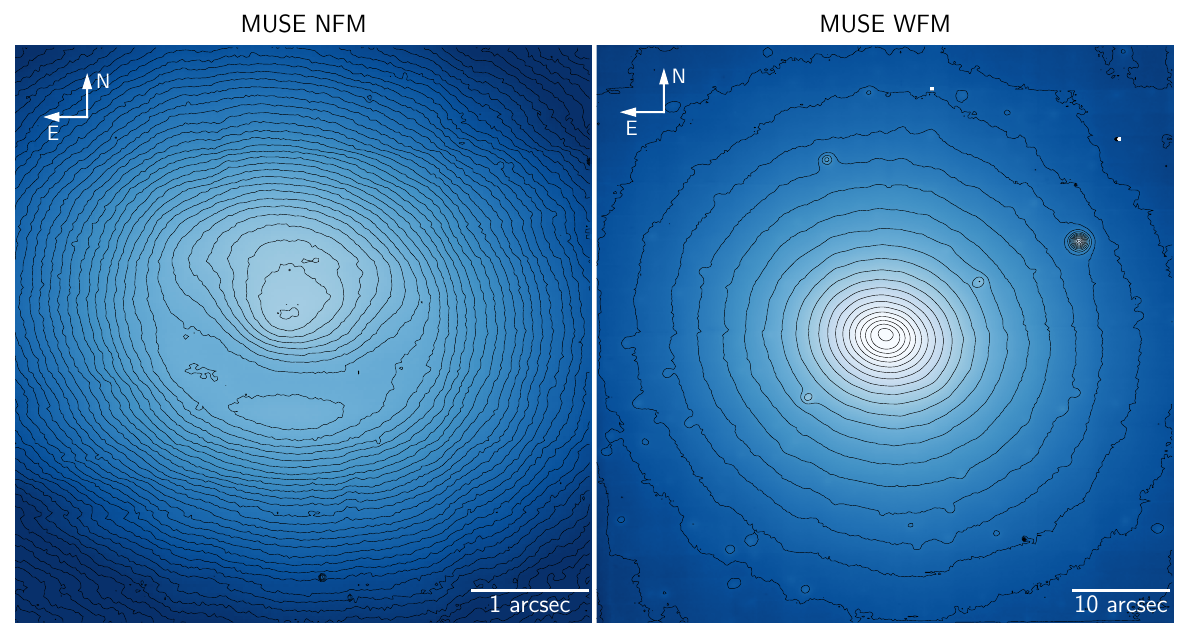}
\caption{Collapsed images (summed over the wavelength axis) of the MUSE NFM (left) and MUSE WFM (right) combined exposures with surface brightness contours (black lines). The contours were chosen to emphasize the circumnuclear dust ring structure in the NFM image (for more detailed images of the dust ring, see \citep{Boizelle2019}) and to highlight the axisymmetry in the WFM data. The NFM image has a $5\arcsec\times5\arcsec$ ($750\rm \ pc \times 750 \ \rm pc$) FoV. The circumnuclear dust ring is evidenced by the drop in surface brightness south of the central isophote of the NFM image. The WFM image shows a $1\arcmin\times1\arcmin$ ($9\rm \ kpc \times 9 \ \rm kpc$) FoV and includes the point source (north and west of NGC 3258) that we measure the WFM PSF from (see Section~\ref{sec: PSF}). Some other small point sources are also visible.
\label{fig: images}}
\end{figure*}

    With such strong correlations between SMBH mass and global galaxy properties, it is essential to understand these relations in detail. SMBH masses form the foundation for our understanding of galaxy formation and evolution and SMBH--galaxy coevolution \citep{Hopkins2007_coevolution, Kormendy&Ho2013}. In order to measure these masses, there are a variety of both direct and indirect techniques available. Some examples of direct methods include gas dynamics, stellar dynamics, and H$\beta$ and X-ray reverberation mapping. Indirect methods include X-ray scaling, fundamental plane (FP), and $M$--$\sigma$ \citep[e.g.][]{Bentz2010, Macchetto1997, Gebhardt2011, Peterson2004, Gliozzi2011, Merloni2003, Gultekin2019, McConnell2013, Walsh2013, Woo2013}. With such a wide array of methods available, it is essential to measure black hole masses in the same galaxy with a variety of techniques to constrain systematic uncertainties. For example, while \citet{Williams2023_methods} found general agreement between the direct methods, they found more than an order-of-magnitude difference in the mass measurement of the SMBH in NGC 4151 when comparing direct mass measurement methods and indirect mass measurement methods. \citet{Gliozzi2024} also conducted a comparison of indirect methods ($M$--$\sigma$, FP, and X-ray scaling methods) with hard X-ray active galactic nuclei (AGN) observations and found that while the FP and X-ray scaling methods are consistent with dynamical mass measurements, $M$--$\sigma$ systematically overestimates the SMBH mass. 
    
    Furthermore, it is important to understand these relations across multiple epochs in cosmic time. Another consequence of the strong SMBH mass--global galaxy property correlations is the likely coevolution of SMBHs with their host galaxies \citep{Kormendy&Ho2013}. However, the evolution, or lack thereof, of SMBH--galaxy scaling relations across cosmic time is still an unanswered question \citep{Peng2006, Alexander2008, Schulze2014, Neeleman2021}. One of the most effective methods to probe this question is in the study of high-redshift AGN. However, illuminating the potential evolution of these scaling relation hinges on understanding (1) the differences and similarities between nearby AGN  and local inactive galaxies and (2) the influence of galaxy morphology and bulge type on AGN \citep{Zhuang2023}.  Another critical pursuit that hinges on the evolution of SMBH mass--galaxy property scaling relations is the constraint on the origin of the gravitational-wave background for which evidence was recently discovered \citep{Agazie2023} and which can be explained by binary SMBHs \citep{Agazie2023_bsmbh} as well as exotic alternatives \citep{2023ApJ...951L..11A}. This relies heavily on understanding SMBH demographics at high redshift (out to $z\sim3$) \citep{Agazie2023_bsmbh}.  \citet{Matt2023}, however, found that SMBH populations predicted using two fundamental scaling relations, $M$--$M_{\mathrm{bulge}}$ and $M$--$\sigma$, are significantly different, suggesting that one or both of these relations may evolve with cosmic time. A fundamental piece needed to answer these questions is a large sample of accurately measured SMBH masses.

\begin{figure}
    \plotone{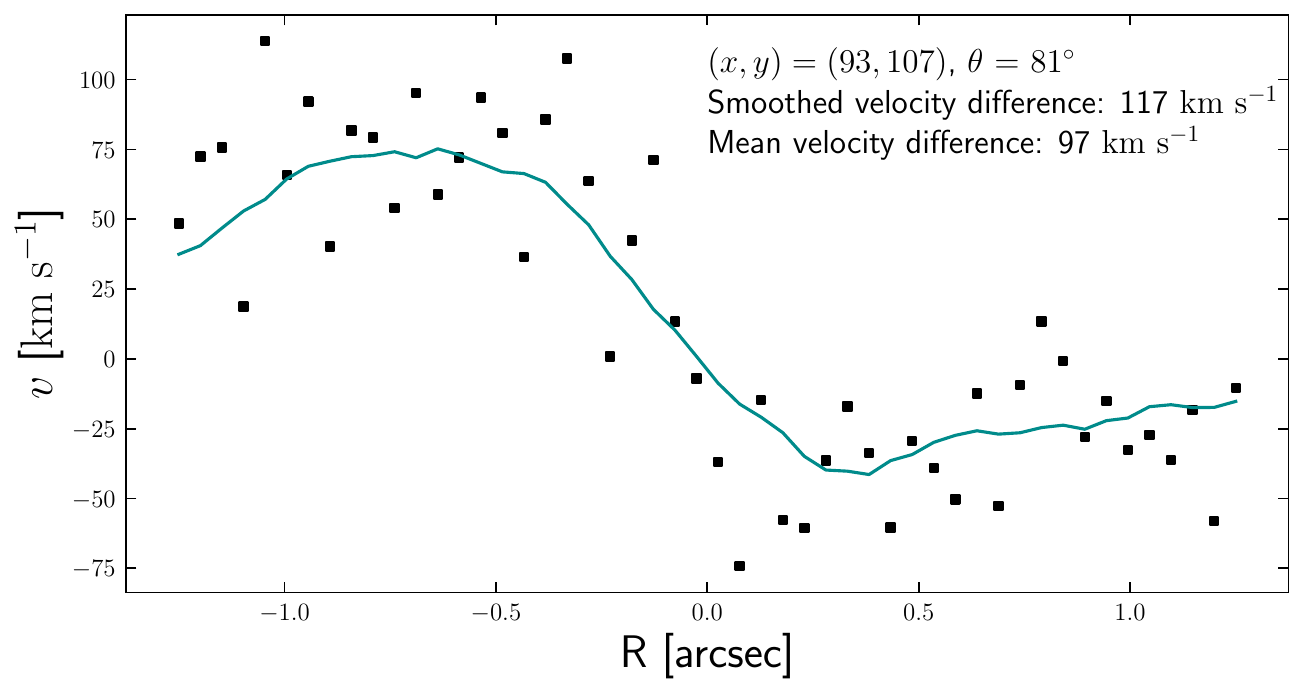}
    \caption{NFM velocities for $50$ $4\times4$ pixel bins across the kinematic center (pixel coordinate $(x,y)=(93,107)$) along the kinematic major axis ($\mathrm{PA} = 81^\circ$). The black points show the velocities computed with pPXF, and the cyan line shows the data smoothed with a Savitzky--Golay filter. Annotated on the plot are the velocity differences across $R=0\arcsec$ using (1) the difference between the minimum and maximum values of the smoothed data and (2) the difference in the mean velocities for radii $|R| > 0\farcs4$.} 
    \label{fig: vel_diff}
\end{figure}

    Direct SMBH mass measurement methods, in particular, are critical as they provide the most accurate constraints on SMBH masses and constitute the underlying basis for all scaling correlations used in estimating the masses of SMBHs within active galaxies. Of the direct methods listed above, stellar dynamics is often considered the standard for SMBH mass measurements \citep{vanderMarel1998, Gebhardt2000}. In this paper, we present a stellar dynamical mass measurement of NGC 3258 which is a nearby bright elliptical (E1) galaxy \citep{deVaucouleurs1991}. Since NGC 3258 is one of two equally bright elliptical galaxies dominating the Antill cluster  \citep{Nakazawa2000} and is relatively close to the Galaxy, it has been studied extensively. \citet{Boizelle2019} conducted a direct measurement of the SMBH mass with CO gas dynamics using data from the Atacama Large Millimeter/submillimeter Array (ALMA). This affords us the opportunity to make a robust comparison between two direct SMBH mass measurement methods and, should they agree, provide strong support for the accuracy of both gas dynamical and stellar dynamical mass measurements. In addition, we will develop a strong constraint on the mass of the SMBH in NGC 3258 that will serve as a benchmark for comparison with indirect mass measurement methods to further constrain systematic uncertainties. 

\begin{figure*}
    \centering
    \includegraphics[width = 1\textwidth]{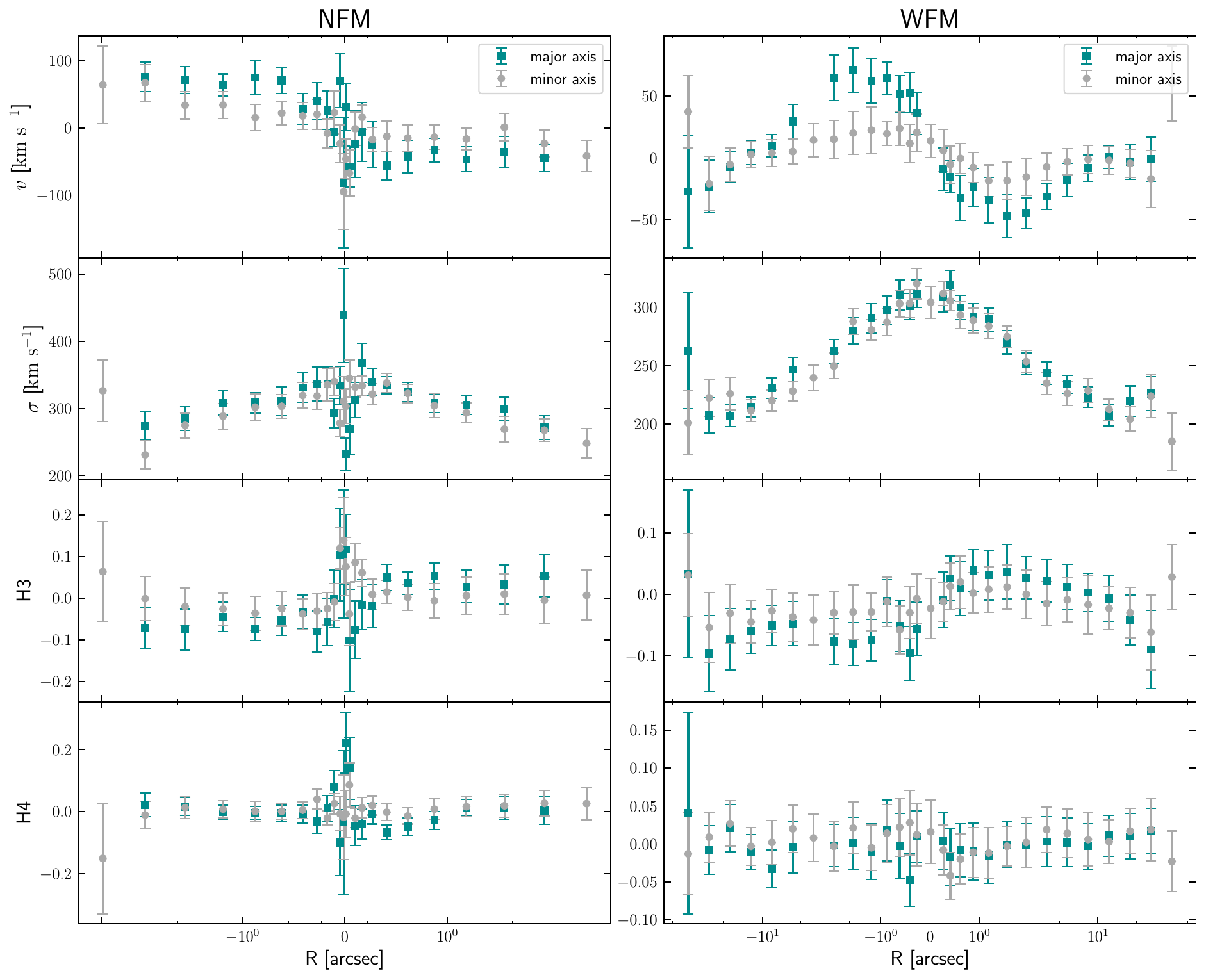}
    \caption{NFM (column 1) and WFM (column 2) radial profiles of the velocities ($v$, row 1) and velocity dispersions ($\sigma$, row 2) calculated from the kinematic extraction and LOSVD fitting and radial profiles of the higher order Gauss-Hermite moments ($h_3$, row 3, and $h_4$, row 4) calculated from the Gauss-Hermite expansion. These data contain points from bins along the kinematic major axis (cyan squares) and the kinematic minor axis (gray circles). The abscissas are presented in a ``asinh'' scaling to show details at small and large scales while also showing similarities and differences on either side of the center. The velocity (and $h_3$) profiles in both the NFM and WFM data show evidence for rotation about the minor axis as well as the major axis. This rotation could be indicative of more complex dynamics (e.g., a triaxial geometry rather than an asymmetric one) or a misclassification of the PA.}
    \label{fig: kinematics}
\end{figure*}

\begin{figure*}[ht]
    \centering
    \includegraphics[height=0.9\textheight]{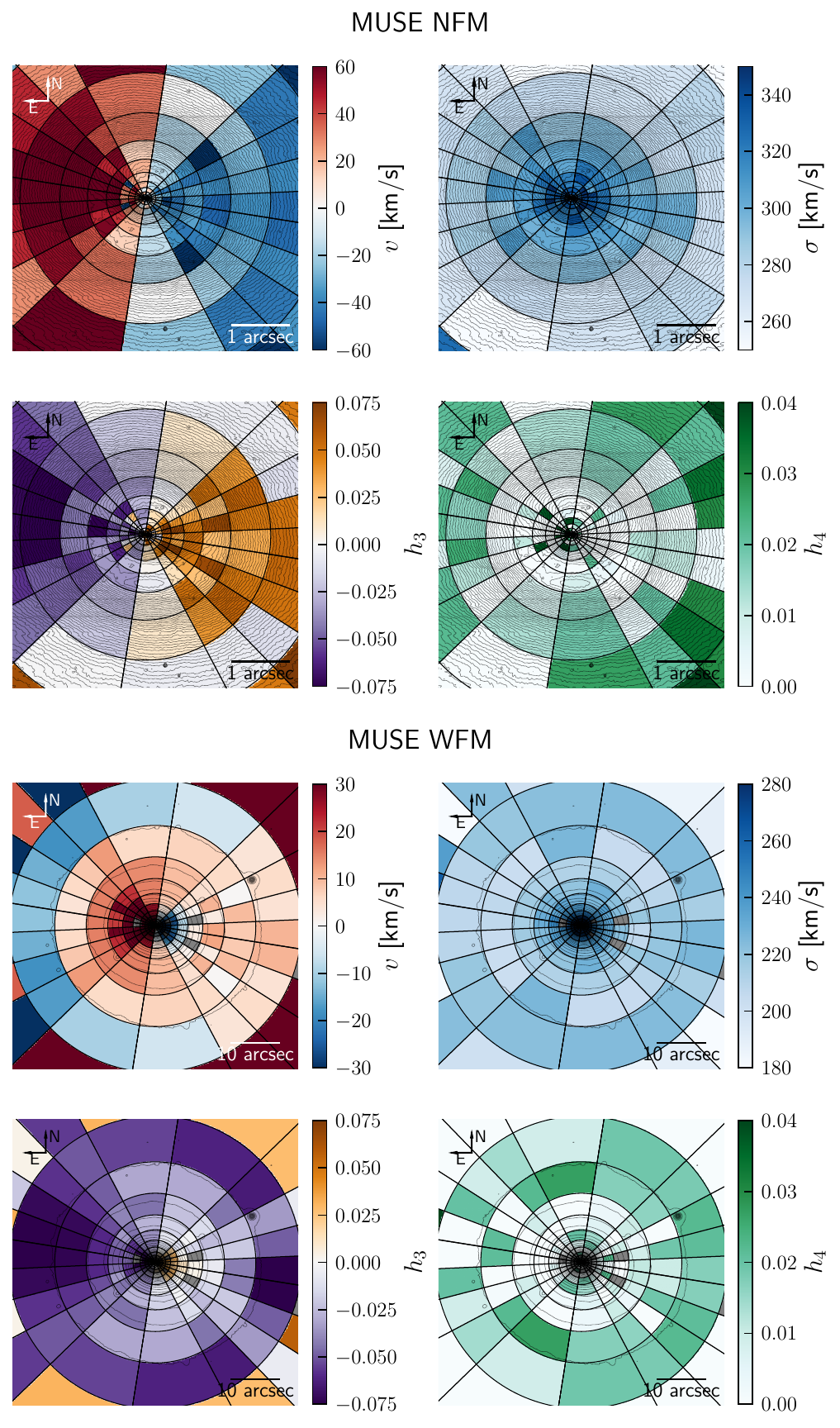}
    \caption{NFM (top four panels) and WFM (bottom four panels) maps of the velocities ($v$, top left in each group) and velocity dispersions ($\sigma$, top right in each group) calculated from the kinematic extraction and LOSVD fitting and maps of the higher-order Gauss--Hermite moments ($h_3$, bottom left in each group; $h_4$, bottom right in each group) calculated from the Gauss--Hermite expansion. Overlaid on each map are surface brightness contours, scale bars corresponding to $1\arcsec$ angular distance ($\sim 150 \rm pc$) for the NFM data and $10\arcsec$ angular distance ($\sim 1.5 \rm kpc$) for the WFM data, and bin edges (black lines). We average data across the kinematic major axis prior to our analysis. For visualization purposes, the data have been reflected across the kinematic major axis.
    \label{fig: gherm}}
\end{figure*} 

    This paper is organized as follows. Section~\ref{sec: observations} outlines our observations and data reduction. Here, we present an overview of the Multi Unit Spectroscopic Explorer (MUSE) and our observations, the MUSE data reduction and processing pipeline, and our postprocessing of the data to make a combined data cube. We also outline our method to determine the kinematic center and kinematic major axis of NGC 3258, and our measurements of the MUSE adaptive optics (AO) point spread functions (PSFs) for the two observational modes: wide-field mode (WFM) and narrow-field mode (NFM). We then present the stellar surface brightness profile, adopted from \citet{Boizelle2019}, that we implement in our modeling. Next, we discuss the models we use to infer the black hole mass ($M$), asymptotic circular velocity ($v_c$), $H$-band mass-to-light ratio ($\Upsilon$), and dark matter halo scale radius ($r_c$) in Section~\ref{sec: model}. We provide details for our modeling method, the Schwarzschild orbit library method, and the results of the modeling. We then present a discussion of this work in Section~\ref{sec: discussion} that explores its implications and outlines future work. Finally, we present a summary of this paper in Section~\ref{sec: summary}. For a robust comparison, we adopt a luminosity distance to NGC 3258 of $31.9 \pm 3.9 \ \rm Mpc$, as computed by \citet{Boizelle2019}. All quantities dependent on distance are scaled to this value.

\begin{figure*}[ht]
    \centering
    \plotone{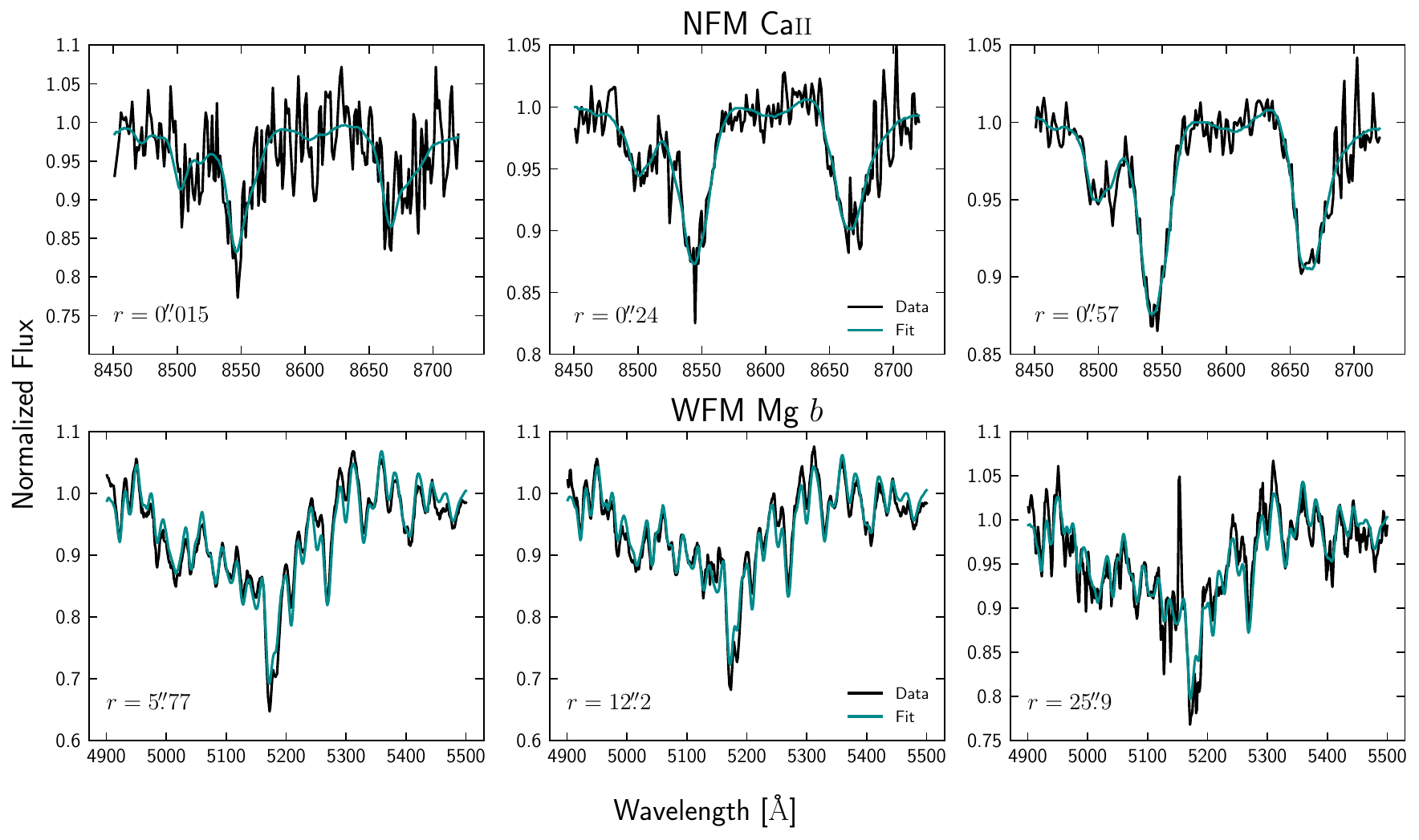}
    \caption{Six example spectral fits for the NFM \CaII triplets for three different radii (as annotated on the plot) near the galactic center (top row) and spectral fits for the WFM \Mgb triplet at radii outside the FoV of the NFM data (bottom row). The data are shown as black lines and the fits, based on the MUSE library of stellar spectra, are shown in cyan. The spectra shown here are a subset of the $20$ bins along the kinematic major axis, which make up a subset of our $200$ total bins. 
    \label{fig: spec_fits}}
\end{figure*}  
    
\section{Observations and Data Reduction} \label{sec: observations}

    We discuss the MUSE and our observations, data reduction, and postprocessing to coadd exposures. We determine the kinematic major axis and kinematic center of NGC 3258 using the penalized pixel fitting (pPXF) method \citep{Cappellari2023, Vazdekis2016} to find the maximum velocity differences across a given center point and a given position angle (PA). We then describe our method for binning our data, extracting the kinematic information from the binned spectra and fitting line-of-sight velocity distributions for both the WFM and NFM data. Next, we outline our measurements of the PSFs for the WFM and NFM data. Finally, we introduce the stellar surface brightness profile that we adopted from \citet{Boizelle2019}. 
    
\begin{figure*}[ht]
    \centering
    \plotone{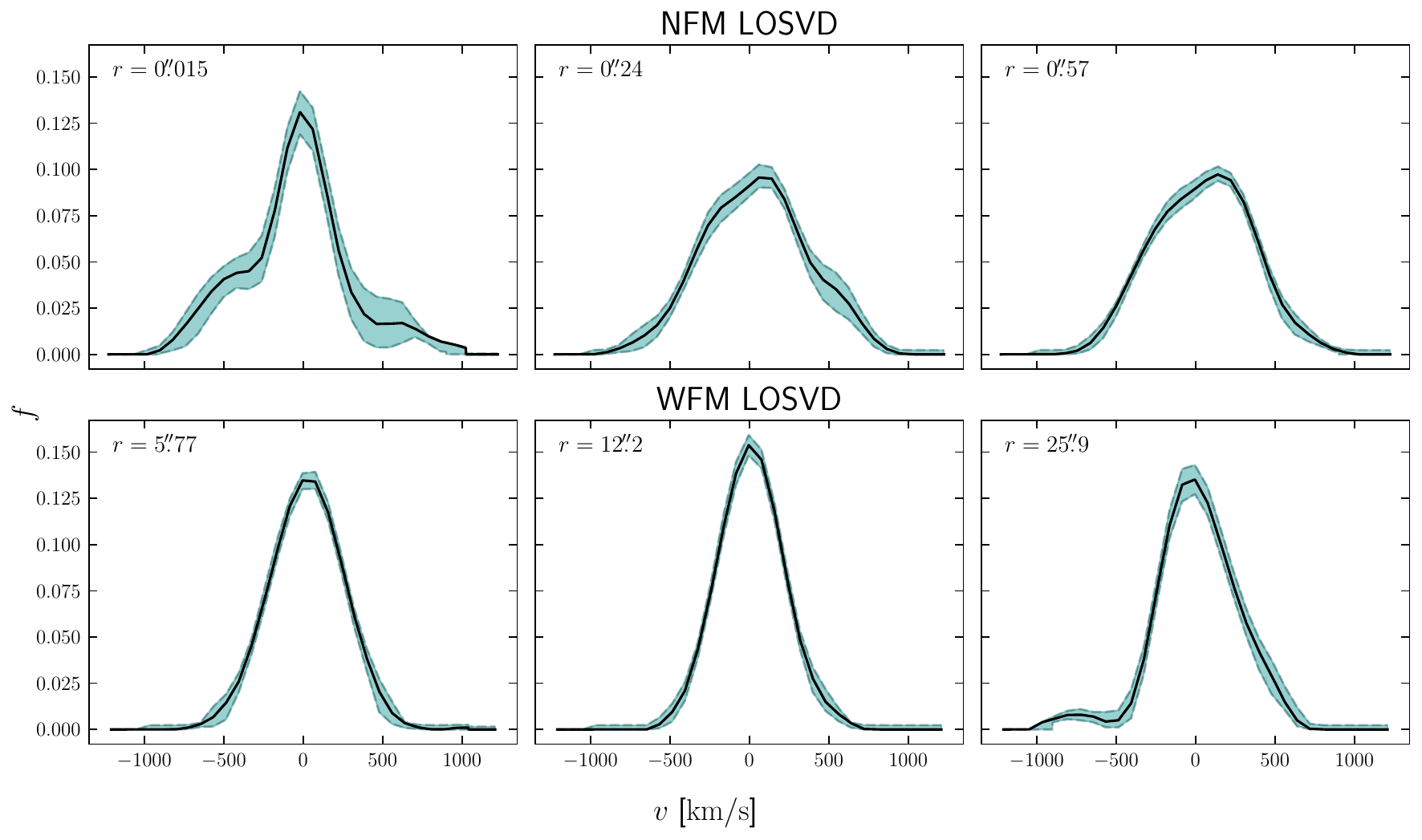}
    \caption{Six example LOSVDs (fraction of stars between velocity $v$ and $v + dv$, $f$, as a function of velocity $v$) for three different radii (as annotated on the plot) in the NFM near the galactic center (top row) and three different radii in the WFM outside the FoV of the NFM data (bottom row). The LOSVD fits are shown as black lines, and uncertainties estimated from MCMC methods are shown as cyan bands. 
    \label{fig: losvds}}
\end{figure*}      

\subsection{MUSE Observations and Data Reduction} \label{sec: muse_data}

    MUSE is a second-generation integral field unit spectrograph (IFU) on the European Southern Observatory (ESO) Very Large Telescope at Cerro Paranal, Chile. Observations of NGC 3258 with MUSE consist of 20 total exposures, taken on the following dates: 2021 February 9, 2021 February 15, 2021 February 22, 2021 March 11, and 2021 April 8 (program ID 105.20K2; PI: Nagar). The observations were taken in both the WFM (2 exposures) and the NFM (18 exposures). The field of view (FoV) of the WFM exposures is approximately $1\arcmin \times 1 \arcmin$ and each of the two exposures has an exposure time of $1680 \ \rm s$. The WFM operates in the wavelength range $4750$--$9350 \ \rm \AA$ and has a spectral resolving power of $R \sim 1770$ ($169\kms$) at $4800  \ \rm \AA$ to $3590$ ($84\kms$) at $9300  \ \rm \AA$. The remaining 18 exposures, each of which has an exposure time of $700 \ \rm s$, were taken in the NFM with the GALACSI AO system ($5\%$ Strehl at $6500 \ \rm \AA$). The NFM FoV is approximately $7\farcs5 \times 7\farcs5$. The NFM operates in the same wavelength range as the WFM and has a similar spectral resolving power of $R \sim 1740$ ($172\kms$) at $4800  \ \rm \AA$ to $3450$ ($69\kms$) at $9300  \ \rm \AA$ \citep{Bacon2010_MUSE}\footnote{Also see the \href{https://www.eso.org/sci/facilities/paranal/instruments/muse/overview.html}{ESO MUSE Overview} and the \href{https://www.eso.org/sci/facilities/develop/ao/sys/galacsi.html}{ESO GALACSI Overview}.}. The mean elevation of the two WFM exposures is $35^\circ$ and the mean elevation of the 18 NFM exposures is $59^\circ$.
    
\begin{figure*}[ht!]
    \centering
    \plotone{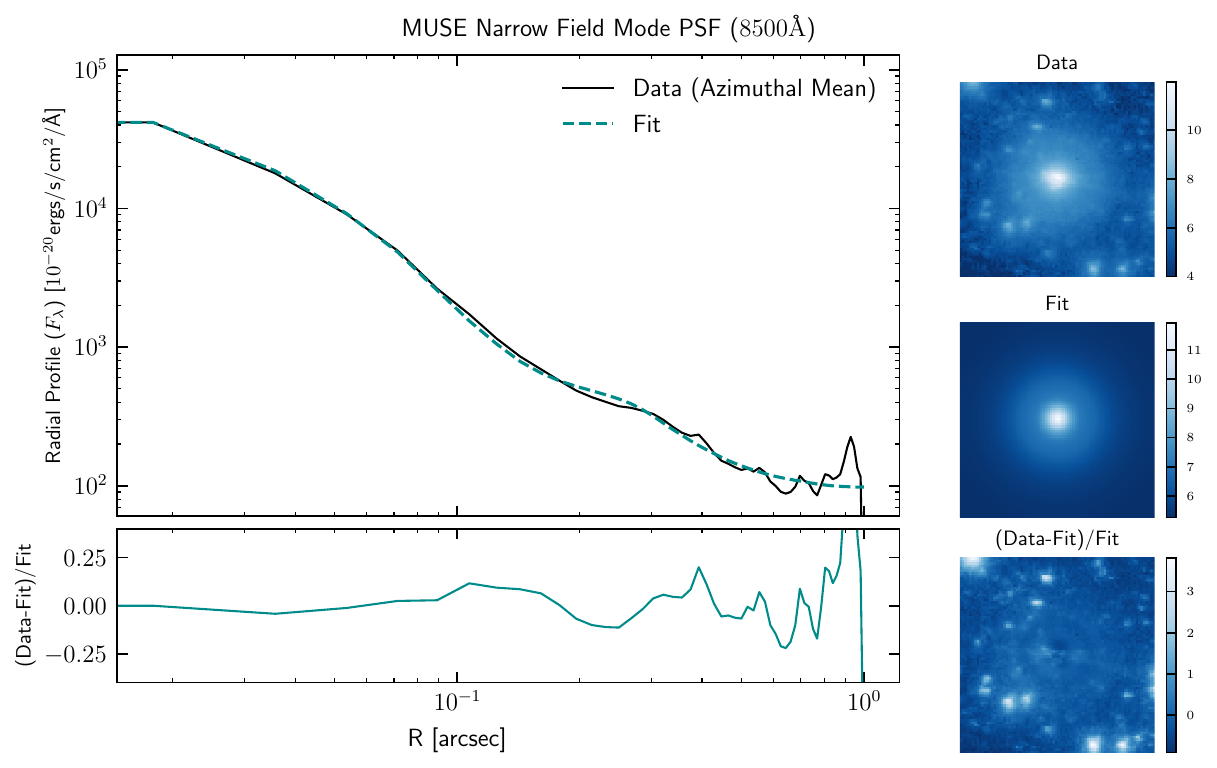}
    \caption{1D (left) and 2D (right) fits to the NFM PSF using the MUSE NFM Psfao model from the MAOPPY library. Data to measure the PSF are composed of the average of five bright point sources from data from the ESO Science Archive Facility at different locations on the sky and with varying exposure times. For the 1D fits (top left), we take the azimuthal mean of the averaged point source and fit the data with the 1D Psfao model. The data are shown as a black solid line, and the fit is shown as a blue dashed line. The residuals for the 1D fit are presented in the bottom left panel. For the 2D fits, we fit the data with the 2D Psfao model. The top right panel shows the data, the center right panel shows the 2D fit, and the bottom right panel shows the 2D residuals. 
    \label{fig: psf_nfm}}
\end{figure*}     

\begin{figure*}[ht]
    \centering
    \plotone{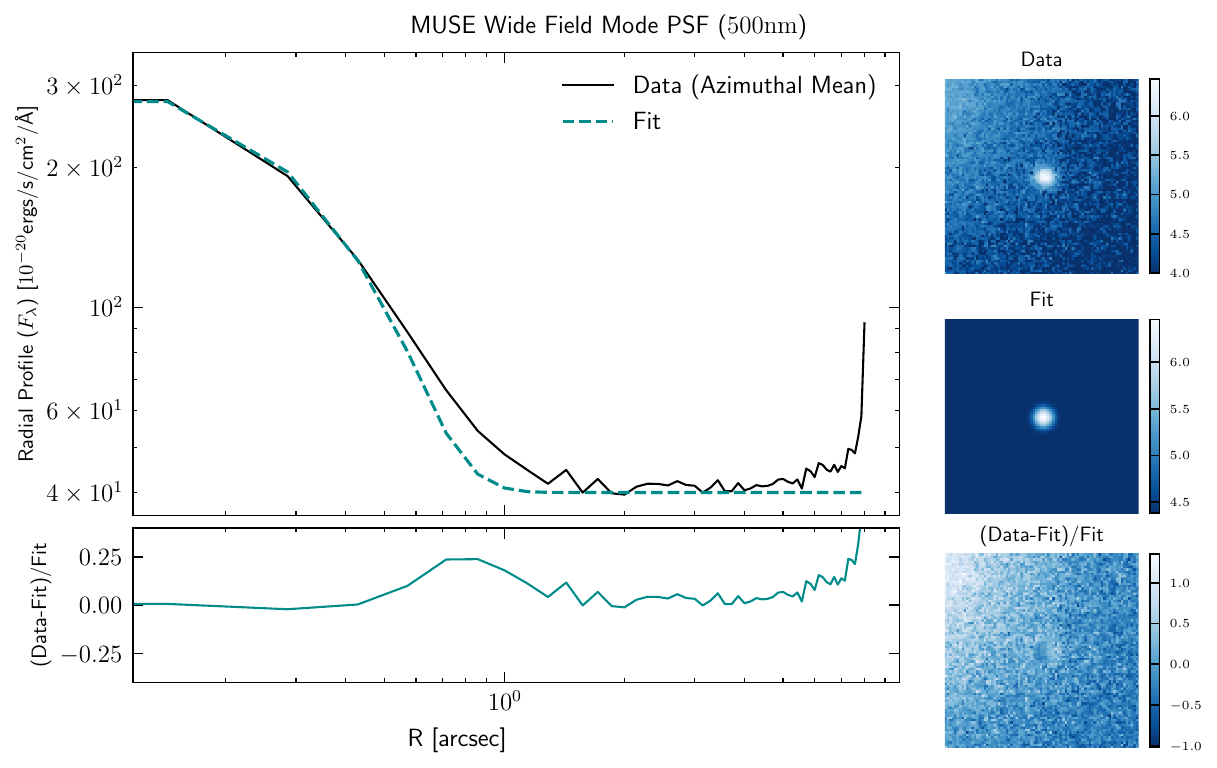}
    \caption{1D (left) and 2D (right) fits to the WFM PSF using a Gaussian function. Data to measure the PSF are composed of 1 bright point source in the WFM FoV (averaged over 2 exposures). For the 1D fits (top left), we take the azimuthal mean of the averaged point source and fit the data with a 1D Gaussian. The data are shown as a black solid line and the fit is shown as a blue dashed line. The residuals for the 1D fit are presented in the bottom left panel. For the 2D fits, we fit a the data with a 2D Gaussian. The top right panel shows the data, the center right panel shows the 2D fit, and the bottom right panel shows the 2D residuals. 
    \label{fig: psf_wfm}}
\end{figure*}

    The MUSE data reduction and processing pipeline calibrates and applies standard data reduction routines, including computing the wavelength solution, to raw science data and then applies on-sky calibrations and combines exposures into data cubes. When combining multiple exposures, the MUSE data reduction pipeline detects point sources with the \texttt{DAOPHOT} detection algorithm \citep{Stetson1987}, calculates coordinate offsets with respect to the first exposure, and shifts and combines exposures into a single data cube \citep{Weilbacher2020}. 

    The exposures in our data set do not contain a sufficient number of point sources to perform the offset calculations, so the MUSE data reduction pipeline returned the processed exposures as 20 individual data cubes. In order to combine our exposures, we performed 2D Gaussian fits to the galactic center (considering only the H$\alpha$ $6563\ \rm \AA$ emission) for each exposure using SciPy's \texttt{optimize.leastsq}. To estimate the uncertainties in the fits, we diagonalized the output covariance matrices and computed the standard error. The resultant errors in the locations of the centroids had a mean of $0\farcs01$ (smaller than the $0\farcs025$ pixel scale of the NFM). Using these fits as a reference point, we calculated the offset between each exposure with respect to the first exposure. We then coadded the spectra from each spaxel and generated a combined data cube for both the NFM and WFM data. We show collapsed images (summed over the wavelength axis) with surface brightness contours for both the WFM and NFM data in Figure~\ref{fig: images}.

\subsection{Determining the Kinematic Center and Major Axis} \label{sec: kin_cent}

    To determine the SMBH mass in NGC 3258, we use the Schwarzschild orbit library method \citep{Schwarzschild1979}, outlined in Section~\ref{sec: schwarz}. This method, in general, assumes either an axisymmetric or a triaxial geometry. Geometries including stellar bars have also been included recently \citep{Valluri2016}. Here, we assume axisymmetry. Therefore, it is important to determine an accurate central point of the galaxy. In addition, our binning scheme uses the kinematic major axis as a reference point. To determine the kinematic center and kinematic major axis of NGC 3258, we looked for the point and the angle that creates a symmetric rotation curve with the greatest amplitude.  To achieve this, we binned the NFM data in $50$ $4\times4$ pixel bins along an arbitrary angle through an arbitrary center pixel. We then computed the velocity of each bin with pPXF \citep{Cappellari2023, Vazdekis2016} in the \CaII triplet region\footnote{Note that we only use pPXF for this calculation as it is a straightforward and quick method of determining the velocity and velocity dispersion. We use our own methods for fitting LOSVDs for modeling, as they do not fit in Gauss--Hermite space, which has drawbacks \citep{2004PASP..116..138C, 2006MNRAS.367....2H}.}. We iterated over all possible PAs between $-110$ and $70$ degrees and over all possible pixel coordinates within a $30\times30$ pixel box centered on the central isophote. We then calculated the velocity difference across the central pixel by (1) smoothing the velocity curves with a Savitzky--Golay filter and taking the difference between the maximum and minimum velocities and (2) calculating the mean of the velocities for $|R| > 0\farcs4$ for the NFM data and $|R| > 15\arcsec$ for the WFM data. We use both of these methods to limit effects by outliers in the data. Both methods 1 and 2 return consistent results for the kinematic center and kinematic major axis PA. The location of the center and the PA of the kinematic major axis is taken to be the angle--coordinate combination that maximizes the velocity difference across the central pixel. 
        
    We found that the kinematic center of the NFM data is located at the pixel coordinates $(x,y)=(93, 107)$ and the kinematic major axis has $\mathrm{PA} = (81\pm1)^\circ$ measured from the north about $(x,y)=(93, 107)$. Our uncertainty estimate for the PA is based on the resolution limit of our binned data for this analysis.  It indicates a minimum level of uncertainty in the kinematic axis, but since we do not marginalize over the PA uncertainty, we simply choose the best-fit value. The kinematic center value is offset from the central isophote ($(x,y)=(96, 110)$) by about 4 pixels or $0\farcs1$. This offset, much smaller than the $\sim1\arcsec$ sphere of influence of the black hole, is likely due to dust obscuration from the circumnuclear dust disk. Figure~\ref{fig: vel_diff} shows the velocity profile that maximizes the difference in velocities for both methods 1 and 2 for the NFM data. To optimize the accuracy of our kinematic analysis and mitigate the effects of excessive scatter, we analyzed the WFM data using a $10\times10$ pixel box centered on the central isophote, organizing the data into $75$ $4\times4$ pixel bins. Recognizing the known PA of the NFM kinematic axis, we focused on a narrower PA range of $-115^\circ$ to $-90^\circ$ to reduce computation time. We determined that the kinematic center in the WFM data cube is located at pixel coordinates $(x, y) = (150, 150)$, which corresponds to the central isophote. Our determination of the kinematic major axis PA is consistent with the PA of $(81\pm1)^\circ$ determined with the NFM data. We find a $4^\circ$ offset from the PA of the circumnuclear CO disk ($\mathrm{PA} = 77^\circ$) as determined by \citet{Boizelle2019}. This offset is unlikely to significantly impact our results as it is much less than the angular size of our bins, $18^\circ$ on average, with the smallest being $\sim11.5^\circ$.

\subsection{Kinematic Extraction and Line-of-Sight Velocity Dispersion Fitting} \label{sec: kin_extract}

    To extract the stellar kinematic information from the MUSE IFU data, we began by binning the data in 20 logarithmically sized radial bins and 20 angular bins equally spaced in $\sin(\theta)$, where $\theta$ is the angle above the major axis, leading to angular bins that are finer closer to the kinematic major axis. Figure ~\ref{fig: kinematics} shows radial profiles of the velocities, velocity dispersions, and $h_3$ and $h_4$ Gauss--Hermite moments and their respective uncertainties estimated from Markov Chain Monte Carlo (MCMC) methods for data along the kinematic major and minor axes. Figure~\ref{fig: gherm}, which shows maps of the velocities, velocity dispersions, and $h_3$ and $h_4$ Gauss--Hermite moments, has our binning scheme overlaid in black lines. Since we assumed axisymmetry, we folded the data across the kinematic major axis and combined the bins. This results in 10 angular bins between  $\mathrm{PA} = -99^\circ$ and $\mathrm{PA} = 81^\circ$. The spectra that fall within these bins are then combined with an unweighted average. 

    We then performed continuum fits and continuum subtraction on the full wavelength range ($4750$--$9350 \ \rm \AA$) for the averaged spectra in each bin. To fit the continuum, we mask out all absorption and emission lines in the spectra by selecting continuum regions by hand and then fit with a fourth-degree polynomial. Next, we fit the \CaII ($8498$, $8542$, and $8662 \rm \AA$) and \Mgb ($5167$, $5173$, and $5184 \ \rm \AA$) triplets in the NFM and WFM data, respectively. To perform these fits, we use the MUSE library of stellar spectra \citep{Ivanov2019}. We use a subset of 10 templates of G and K giants, whose spectra contain strong \CaII and \Mgb features and dominate the galactic spectra of early-type galaxies. These lines are the most common tracers for galaxy kinematics. \CaII, in particular, is advantageous due to its insensitivity to extinction \citep{Silge2003}. We use \Mgb in the WFM data due a to larger uncertainty in \CaII fits. Since the WFM data are primarily used to constrain kinematics outside the NFM FoV, dust obscuration will not significantly impact the spectral fits. Our template stars have the following spectral types: G0, G5Ia, G8Iab, G5IIIa, K0, K0III, K2III, K3IIp, K3.5III, and K5III. Figure~\ref{fig: spec_fits} shows \CaII fits for three bins along the kinematic major axis near the galactic center using the NFM data and \Mgb fits for three bins at larger radii using the WFM data. 

    Finally, to obtain the kinematic information, we follow the method of \citet{Gebhardt2000}. We deconvolve the galactic spectra within each bin using the MUSE K and G giant spectral templates using a maximum penalized likelihood (MPL) estimate similar to the techniques employed by \citet{Saha1994} and \citet{Merritt1997}. The MPL estimate chooses an initial velocity profile in each bin, convolves the profile with a weighted average of the MUSE K and G giant templates, and calculates the residuals from the galaxy spectra in each bin. The program then iteratively varies the velocity profile parameters and template weights and computes and minimizes the $\chi^2$ to return a best-fit, nonparametric, line-of-sight velocity dispersion (LOSVD) for each bin. We smooth the resultant LOSVDs by adding a penalty function to the $\chi^2$. This fit is performed in LOSVD space. Finally, we estimate the uncertainties on the LOSVDs using an MCMC approach. We also provide the $h_3$ and $h_4$ Gauss--Hermite moments to communicate these results and summarize the data. Figure~\ref{fig: gherm} shows maps of $v$ (top left), $\sigma$ (top right), and $h_3$ (bottom left) and $h_4$ (bottom right) for the NFM data. The data have been reflected across the kinematic major axis to create a full map of the galaxy. For a more detailed discussion on this method, see \citet{Gebhardt2000}. Figure~\ref{fig: losvds} shows LOSVDs for three bins along the kinematic major axis near the galactic center using the NFM data and LOSVDs for three bins at larger radii using the WFM data.

\subsection{WFM and NFM PSFs} \label{sec: PSF}

    While ground-based astronomy enjoys the advantage of telescopes with very large collecting areas, a major drawback to data collection is atmospheric turbulence, which causes aberration in the incoming wave fronts as they pass through the atmosphere \citep{Roddier1981}. To correct for these aberrations and move toward the diffraction-limited regime, modern telescopes employ AO that use wave-front sensors and deformable mirrors to correct for the effects of atmospheric turbulence \citep{Roddier1999}. AO correction, however, is limited by several factors, including sensor noise, the number of actuators, and loop delay \citep{Martin2017,Rigaut1998}. Since atmospheric turbulence cannot be corrected for completely, the resultant PSF of AO systems is complex. 

    It is critical to understand the PSF to accurately deconvolve long exposures. Since this work relies on fine spatial resolution, deconvolution is a necessary step to extract accurate information, especially on the scale of the SMBHs sphere of influence \citep{Starck2002}. Unfortunately, our NFM data do not contain point sources from which to measure the PSF directly. To measure the NFM PSF, we took the average of bright point sources from 5 exposures\footnote{The data were obtained from the ESO Science Archive Facility \citep{esodata}.} from the ESO Science Archive Facility with various locations on the sky and exposure times ranging from $80$ to $1200\ \rm s$ at a wavelength of $8500\ \rm \AA$. This wavelength was chosen to maximize PSF accuracy at the \CaII triplet. The mean elevation of these exposures is $56^\circ$. In addition, the MUSE NFM PSF is strongly dependent on seeing and GALACSI tilt-tip star (TTS) guide star magnitude \citep{Strobele2012}. The average seeing for our observations is 0.83\arcsec, and the average seeing for the point-source exposures is 0.39\arcsec. Therefore, the PSF measured from the point sources has a higher Strehl ratio than the PSF of the NGC 3258 observations. The variation of the PSF across the five point-source exposures is minimal due to relatively consistent seeing (with a standard deviation of 0.08\arcsec). Variation in the PSF across the NCG 3258 exposures is more significant, as the seeing ranges from 0.49\arcsec to 1.37\arcsec with a standard deviation of 0.28\arcsec. The magnitude of the GALACSI TTS guide star in the point-source observations ranges from 8.1 to 10.1 in the $G_{RP}$ band with a mean of 10.0 and standard deviation of 1.0 mag. The NGC 3258 observations, which all share the same guide star, have a GALACSI TTS guide star magnitude of 11.5. This will result in a slightly lower Strehl ratio than in the point-source observations. However, since the NFM PSF FWHM is much smaller the SMBH's sphere of influence ($\sim1\farcs14$), we do not expect small variations in the PSF to significantly impact our results.
    
    Using the averaged point source, we fit the PSF in 1D and 2D using the MUSE NFM AO-corrected PSF (Psfao) model from the Modelization of the Adaptive Optics Psf in PYthon (MAOPPY) library. In general, AO PSFs can be well approximated by Moffat functions \citep{Trujillo2001,Moffat1969}, but we used the Psfao model because of its flexibility.  A full mathematical description of the Psfao model is presented in \citet{Fetick2019} and extensions to other instruments, such as Keck AO, SOUL at Large Binocular Telescope, CANARY, and the William Herschel Telescope (WHT) and GEMS/GSAOI at GEMINI, are presented in \citet{Beltramo2020}. The main features of this model are that (1)  telescopic pupil diffraction and telescope phase aberrations are taken into account, (2) the model describes the two-component AO PSF, which is composed of a Gaussian-like peak and a turbulent halo, (3) the model supports elliptical asymmetry, (4) the turbulent halo can be estimated outside the telescope FoV, and (5) the model can manage undersampled PSFs \citep{Fetick2019}. This model has also been used with success in several other works \citep[e.g.,][]{Fetick2020, Gottgens2021, Massari2020, Schreiber2020}. Figure~\ref{fig: psf_nfm} shows the 1D Psfao fit for an azimuthal mean of the averaged point source (top left) and the residuals of the fit (bottom left). Also shown in Figure~\ref{fig: psf_nfm} is the averaged point source (top right), the 2D Psfao fit (center right), and the residuals (bottom right). The resultant PSF has an FWHM of $0.035\arcsec$ The parameters for this fit are provided in Table~\ref{tab: nfm_psf}. 

\begin{deluxetable}{llll}
    \tablecaption{NFM PSF Psfao fit parameters} 
    \tablehead{\colhead{Parameter} & 
    \colhead{Definition} &
    \colhead{Fit Value} &
    \colhead{Unit}}
\startdata
    $r_0$     & Fried parameter   & 0.266     & $\rm cm$          \\ 
    $A$       & AO variance       & 3.71      & $\rm rad^2$       \\ 
    $C$       & AO area constant  & 0.0128    & $\rm rad^2 \ m^2$ \\ 
    $\alpha$  & Moffat transition & 0.143     & $\rm m^{-1}$      \\ 
    $\beta$   & Moffat power law  & 2.18      & \dots             \\ 
    $r$       & Axis ratio        & 1.00 (fixed)     & \dots                
\enddata
    \tablecomments{Definitions of symbols in PSF model used for NFM data with best-fit parameter values. \label{tab: nfm_psf}} 
\end{deluxetable}

    The WFM data contain a single bright point source, which allows us to measure the PSF directly from our data set by applying the same method as used with the NFM data. Since our spatial bins in the outer regions of the galaxy are much larger than the size of the PSF, we take the simpler approach of fitting 1D and 2D Gaussians to the point source in order to measure the PSF. The resultant fit has an FWHM of $0.706\arcsec$. Figure~\ref{fig: psf_wfm} shows the 1D Gaussian fit for an azimuthal mean of the point source from the WMF exposures (top left) and the residuals of the fit (bottom right). Also shown in Figure~\ref{fig: psf_wfm} is the point source from the WFM data (top right), the 2D Gaussian fit (center right), and the residuals (bottom right). The 1D residuals of the Gaussian fit reach approximately $25\%$ near the wings of the PSF. Since our spatial bins are much larger than the FWHM of the WFM PSF, we do not expect this to significantly impact the results of our measurements.

\subsection{Stellar Surface Brightness Profile} \label{sec: surface_brightness}

    As a central point of this work is to present a robust comparison between the SMBH mass measured with gas dynamics (from \citealp{Boizelle2019}) and the SMBH mass measured with stellar dynamics (this work), we adopt the stellar surface brightness profile from \citet{Boizelle2019} to make this comparison as direct as possible. Measurements to construct the surface brightness profile were taken  in the $H$ and $V$ bands with the \textit{Hubble Space Telescope} Wide Field Camera 3 in the near-infrared mode, program ID GO-14920, and \textit{HST} Advanced Camera for Surveys Wide Field Channel, obtained from the \textit{HST} archive. Taking measurements along the major axis at large radii, they found a background level of $H=20.8 \ \rm mag \ arcsec^{-2}$ and a color of $V-H=2.40 \ \rm mag$, which was used to align the $H$ and $V$ band profiles. They then applied a correction for Galactic reddening ($A_H = 0.041 \ \rm mag$; \citealp{Schlafly2011}). They modeled the stellar surface brightness profile with the multi-Gaussian expansion (MGE) model from \citet{Emsellem1994} and \citet{Cappellari2002}. Since NGC 3258 has a circumnuclear dust ring, they masked the dust-obscured regions from $R \sim 0\farcs15$ to $0\farcs8$ (as well as contaminating galaxies and foreground stars). 

\begin{figure}[t!]
    \plotone{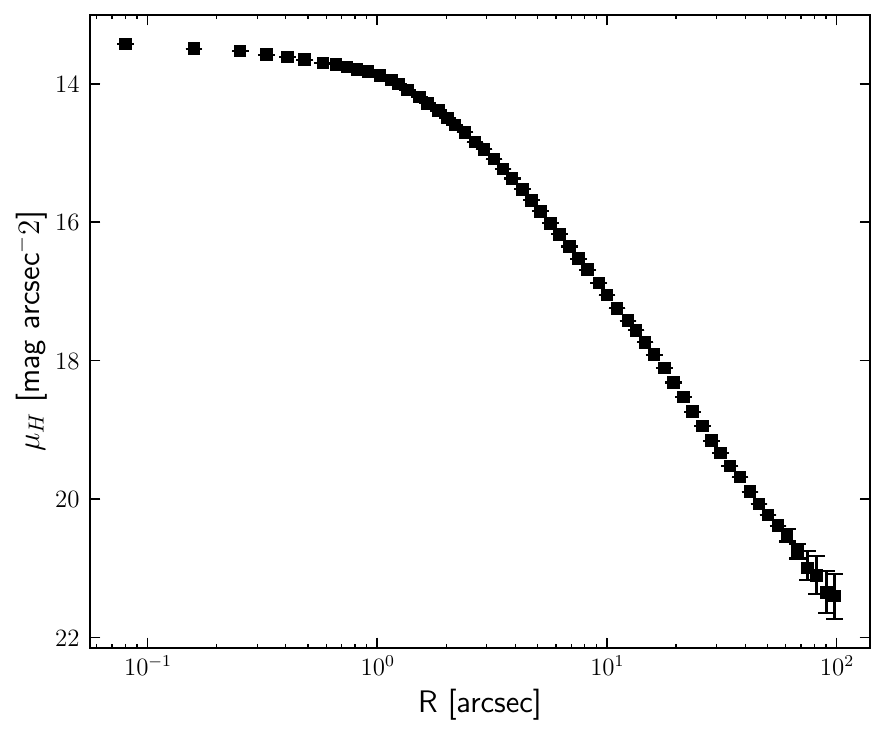}
    \caption{Stellar surface brightness profile adopted from \citet{Boizelle2019} as discussed in section~\ref{sec: surface_brightness}) measured along the major axis from \textit{HST} $H$-band images. The stellar surface brightness profile displays a flattening in the slope at approximately $1\arcsec$, indicative of a cored galaxy profile with a break radius of $\sim 1\arcsec$. The circumnuclear dust ring and foreground galaxies and stars have been masked out. 
    \label{fig: sb}}
\end{figure}

    The resultant MGE model consists of $14$ concentric, elliptical Gaussians and returns a total $H$-band luminosity of $L_{H} = 1.1 \times 10^{11} \ \rm L_{\odot}$ (taken within the central $300\arcsec$ or $45.5 \ \rm kpc$). For a more detailed description of the determination of the surface brightness profile and a table of the MGE best-fit parameters, see \citet{Boizelle2019}. The $H$-band surface brightness as a function of radius (in arcseconds) is shown in Figure~\ref{fig: sb}. The surface brightness profile has a flattening in the slope beginning at radii less than $\sim 1\arcsec$, thus displaying a clear cored galaxy profile \citep{Lauer2007, Rusli2013, Trujillo2004} with a break radius of approximately $1\arcsec$.

\section{Modeling} \label{sec: model} 

    In this section, we discuss our implementation of the Schwarzschild orbit library method \citep{Schwarzschild1979} and the results of our modeling: estimates for the black hole mass ($M$), stellar $H$-band mass-to-light ratio ($\Upsilon$), asymptotic circular velocity ($v_c$), and the dark matter halo scale radius ($r_c$). 
    
\subsection{Schwarzschild Orbit Library Kinematic Models} \label{sec: schwarz}

    We use axisymmetric, three-integral orbit-based models, based on the Schwarzschild orbit library method \citep{Schwarzschild1979}, to compute the SMBH mass ($M$), the mass-to-light ratio ($\Upsilon$), the asymptotic circular velocity ($v_c$), and the dark matter halo scale radius ($r_c$) following the method described in \citet{Gebhardt2000, Gebhardt2003} and, in detail, in \citet{Siopis2009}. The Schwarzschild orbit library method is a general stellar dynamical equilibrium model of a self-gravitating system that is used to infer central black hole masses. 
    
    Our workflow, in general, is as follows. We first adopted the stellar surface brightness profile from \citet{Boizelle2019} (see Section~\ref{sec: surface_brightness} of this work for a description of how they generated this profile) and converted it to a stellar luminosity density distribution by deprojecting the surface brightness profile, assuming axisymmetry, and assuming an inclination equal to that of the circumnuclear dust disk ($i\sim48^\circ$; \citealt{Boizelle2019}). We then converted the luminosity density to a mass density with an unknown but parameterized $H$-band mass-to-light ratio ($\Upsilon$). Next, we determined the stellar gravitational potential by solving Poisson's equation. To the stellar gravitational potential, we added a point mass (black hole with mass $M$) and a dark matter halo parameterized as a cored logarithmic profile, which corresponds to a dark matter density profile of
    \begin{equation} \label{eq: dm_density}
        \rho_{\mathrm{DM}}(r)=\frac{v_c}{4 \pi G} \frac{3r_c^2 + r^2}{(r_c^2 + r^2)^2},
    \end{equation}
    where $r_c$ is the core radius ($\rho_{\mathrm{DM}} \rightarrow \mathrm{constant}$ for $r \ll r_c$) and $v_c$ is the asymptotic circular speed as $r \rightarrow \infty$ \citep{Persic1996}.
    This results in a parameter space defined by the central black hole mass, stellar $H$-band mass-to-light ratio, asymptotic circular velocity, and dark matter halo scale radius. 
    
\begin{figure*}[t!]
    \plotone{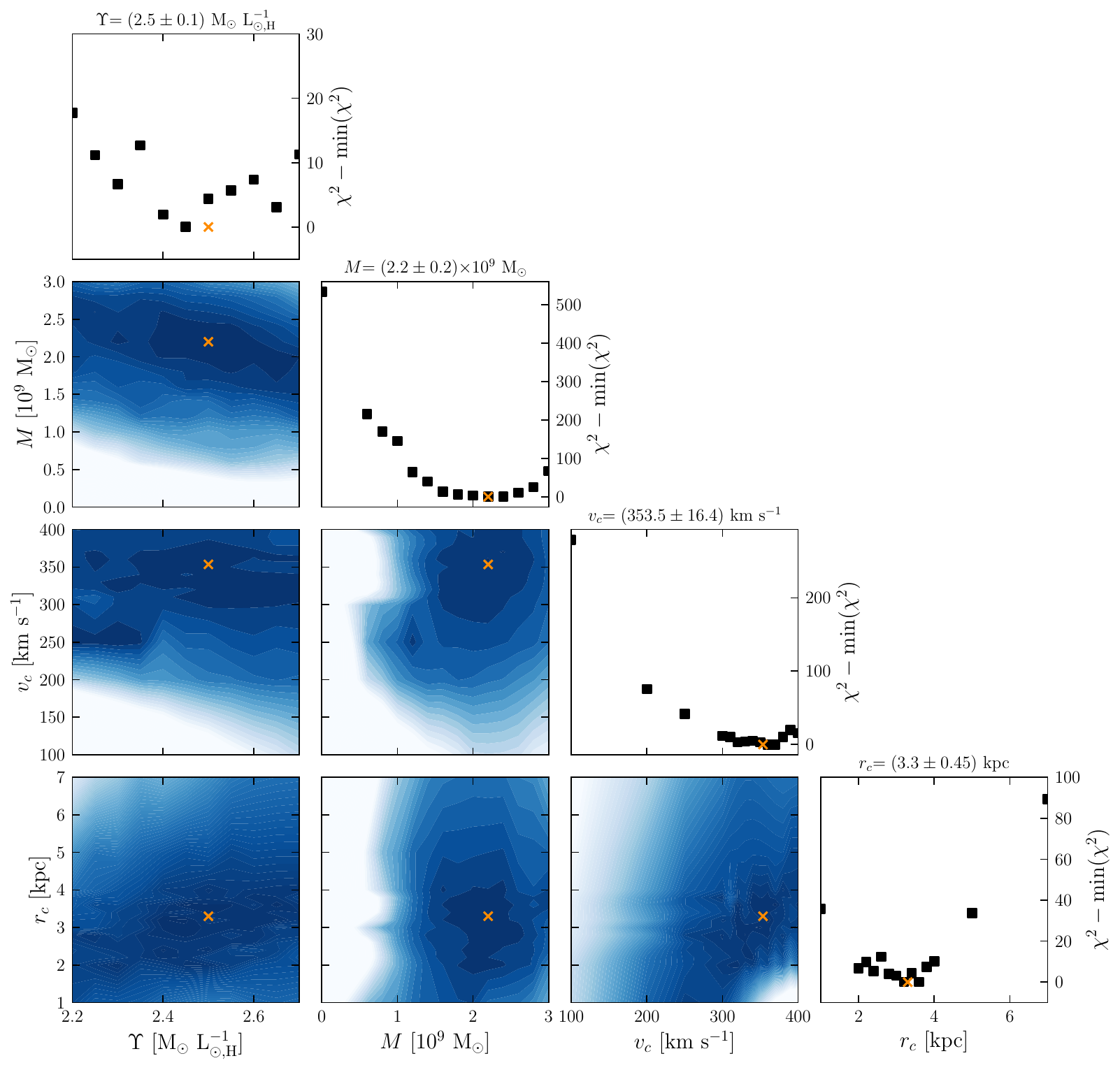}
    \caption{Model output $\chi^2$ contours for six combinations of the three-integral axisymmetric Schwarzschild orbit-library-based modeling method parameters: SMBH mass ($M$), $H$-band mass-to-light ratio ($\Upsilon$), asymptotic circular velocity ($v_c$), and dark matter halo scale radius (off-diagonal plots). We also present marginalized $\chi^2$ values as a function of each parameter on the diagonal plots. In each panel, out best-fit values are shown as orange crosses.
    \label{fig: chi2}}
\end{figure*}

    We then, for a given set of parameters, calculated a library of many ($10^3$--$10^4$) orbits of representative stars. From this library of orbits, the time spent by representative stars and the line-of-sight velocity for each bin are recorded. Using the orbits of the representative stars, we determine the best set of nonnegative orbital weights so that the model LOSVDs reproduce the observed LOSVDs (as determined in Section~\ref{sec: kin_extract}) and the sum of the light in each bin, which is determined by the time the representative stars spend in that bin, reproduces the observed surface brightness profiles. For this comparison, we convolve the resultant models with the PSFs for the WFM and NFM data (see Section~\ref{sec: PSF}).
    
    To narrow the parameter space and find a global minimum, we conducted a set of runs with a coarse grid over the following parameter space: $2.2 \leq \Upsilon \leq 2.7 \MsunLsun$, $0 \leq M \leq 3.0 \times 10^9 \Msun$, $100 \leq v_c \leq 400 \ \rm km \ s^{-1}$, and $1.0 \leq r_c \leq 7.0 \ \rm kpc$. We use $\chi^2$ statistics to determine the goodness of the fit and find estimates for the best-fit black hole mass, stellar $H$-band mass-to-light ratio, asymptotic circular velocity, and dark matter halo scale radius that minimize $\chi^2$. After obtaining an initial set of parameters, we iterated over a fine grid near the best fit values to determine more accurate parameters and best determine the uncertainties in each parameter. 

\subsection{Modeling Results} \label{sec: schwarz_results}  

\begin{figure}[t]
    \plotone{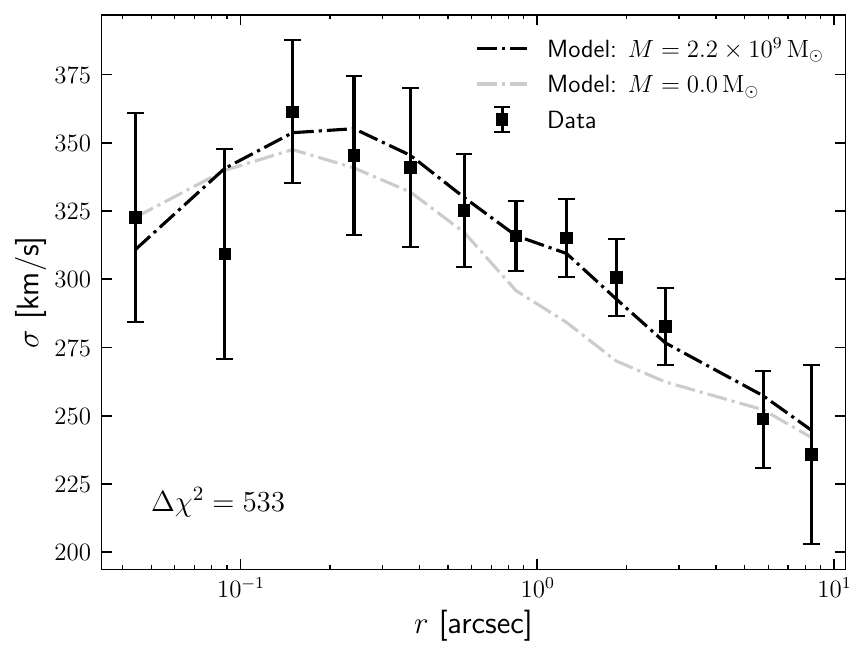}
    \caption{Velocity dispersion ($\sigma$) as a function of radius for radial bins along the kinematic major axis. Our data are plotted as black squares. Overlaid on the plot are the overall best-fit model for $M = 2.2 \times 10^9 \Msun$ (black dotted-dashed line) and the best-fit $M = 0.0 \Msun$ (gray dashed line). The model corresponding to $M = 2.2 \times 10^9 \Msun$ is our best-fit model from the three-integral axisymmetric Schwarzschild orbit-library-based modeling method. The zero-SMBH-mass model is plotted as comparison to show the strong confidence of the presence of an SMBH in NGC 3258, as the $M=0$ curve is nearly everywhere worse than our best-fit model. The difference in $\chi^2$ between these two models is $\Delta\chi^2 = 533$. 
    \label{fig: model_comp}}
\end{figure}

    We report best-fit values and $1\sigma$, $2\sigma$, and $3\sigma$ confidence intervals for (1) black hole mass as $M = (2.2 \pm 0.2 \ (0.3,\  0.5))\times10^9 \Msun$, (2) $H$-band mass-to-light ratio as $\Upsilon = 2.5 \pm 0.1 \ (0.15, \ 0.2) \MsunLsun$, (3) asymptotic circular velocity as $v_c = 353.5 \pm 16.4 \ (30.3, \ 38.3)\ \rm km \ s^{-1}$, and (4) dark matter halo scale radius as $r_c = 3.3 \pm 0.45 \ (0.85, \ 1.5) \ \rm kpc$. Figure~\ref{fig: chi2} shows $\chi^2$ contours (off-diagonal plots) for six combinations of the model parameters and marginalized $\chi^2$ values as a function of each parameter (plots on the diagonal). Overlaid on each are the best-fit values for each parameter, shown as orange crosses. 
    Figure~\ref{fig: model_comp} shows a comparison of velocity dispersion profiles ($\sigma(r)$) between two different models: the overall best-fit model ($M = 2.2\times10^9 \Msun$; black doted--dashed) and the best-fit no-black-hole model ($M = 0.0 \Msun$; gray dashed). 
    Also annotated on this plot is the difference in $\chi^2$ ($=533$) between the two models, which indicates the presence of an SMBH in NGC 3258 with a high degree of certainty.      

    Using the output values of the black hole mass, stellar $H$-band mass-to-light ratio, asymptotic circular velocity, and dark matter halo scale radius and the $\chi^2$ for each model, we compute the best-fit values for each parameter as well as their $1\sigma$, $2\sigma$, and $3\sigma$ confidence intervals corresponding to $\Delta \chi^2$ values of $1$, $3$, and $9$, respectively. Based on tests using models with very fine spacing of parameters close to the best fit, we estimate an RMS noise in the $\chi^2$ of approximately 4.  This is a result of the discrete nature of orbit-based models as discussed in \citet{Vasiliev2020}.  We use this value to smooth the results and calculate our statistical uncertainties. 
    
\begin{figure}[t!]
    \plotone{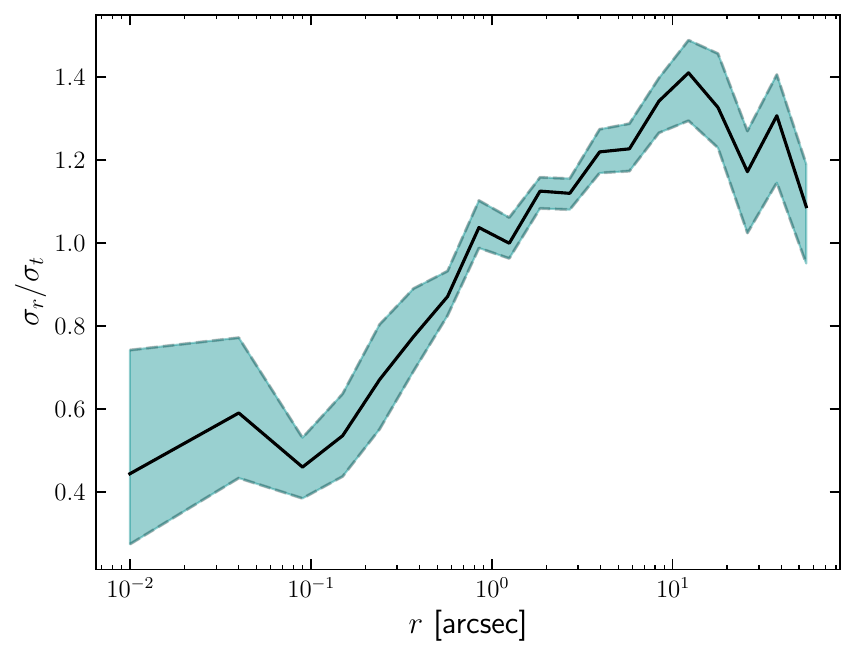}
    \caption{Ratio of the radial velocity dispersion ($\sigma_r$) to the tangential velocity dispersion ($\sigma_t$) as a function of radius (in arcsec). The tangential velocity dispersion ($\sigma_t$) is defined as $\sigma^2_t \equiv 0.5(\sigma^2_\theta + \sigma^2_\phi)$, where $\sigma_\phi$ is the second moment of the azimuthal velocity relative to the systemic velocity and $\sigma_\theta$ is the velocity dispersion in the zenith angle direction, so that $\sigma_r/\sigma_t = 1$ for an isotropic distribution. Tangential motions dominate in the inner $\sim 1\arcsec$ (approximately equal to the break-radius evident in the surface brightness profile in Figure~\ref{fig: sb}), after which radial motions take over the profile.  
    \label{fig: sigr_sigt}}
\end{figure}

    In addition, using the model outputs, we compute the ratio of the radial velocity dispersion to the tangential velocity dispersion ($\sigma_r/\sigma_t$) as a function of galactic radius to quantify the anisotropy. The tangential velocity dispersion ($\sigma_t$) is defined as $\sigma^2_t \equiv 0.5(\sigma^2_\theta + \sigma^2_\phi)$, where $\sigma_\phi$ and $\sigma_\theta$ are the velocity dispersions in the standard spherical coordinate directions, so that $\sigma_r/\sigma_t = 1$ for an isotropic distribution. Figure~\ref{fig: sigr_sigt} shows this profile (black) and its uncertainties (cyan). The anisotropy crosses unity at approximately $1\arcsec$ and is tangentially biased in the inner arcsecond. This bias may be explained by (1) the circumnuclear disk structure, which has a similar spatial extent, and/or (2) a core-scouring event from a coalesced SMBH binary \citep{Harris2024, Ravindranath2002, Thomas2014}. The latter is also evidenced by the cored galaxy profile seen in Figure~\ref{fig: sb} and discussed in Section~\ref{sec: surface_brightness}, which shows a break radius of approximately $1\arcsec$.
    
\section{Discussion} \label{sec: discussion}

    Here, we compare our stellar dynamical mass measurement of the SMBH in NGC 3258 to values obtained from \citet{Boizelle2019} and the $M$--$\sigma$ and $M$--$L$ relations from \citet{Kormendy&Ho2013}. We also discuss the difference between our measured mass-to-light ratio and that measured by \citet{Boizelle2019}. Finally, we explore the consequences of the cored galaxy profile for our  $M$--$\sigma$ estimates. 

\subsection{Comparison to the \texorpdfstring{\citet{Boizelle2019}}{Boizelle et al.\ (209)} Mass and the \texorpdfstring{$M$--$\sigma$ and $M$--$L$ Relations}{M–sigma and M–L Relations}} \label{sec: M-sigma_M-L}

    The results from the three-integral axisymmetric Schwarzschild orbit-library-based modeling are as follows: $M = (2.2 \pm 0.2)\times10^9 \Msun$, $\Upsilon = 2.5 \pm 0.1 \MsunLsun$, $v_c = 353.5 \pm 16.4 \ \rm km \ s^{-1}$, and $r_c = 3.3 \pm 0.45 \ \rm kpc$. The modeling due to \citet{Boizelle2019} covers an impressive amount of data and model assumptions in order to quantify systematic uncertainties.  They model both Cycle 2 and Cycle 4 ALMA data, consider various degrees of extinction correction to the light profile, try both flat and ``tilted ring'' disk structures, and assume uniform and Gaussian turbulent velocity profiles.  Their preferred model is fitted to Cycle 4 data using a titled ring model with Gaussian turbulent velocity profile and model of extended mass that is only constrained by the ALMA data and allowed to vary with radius as long as the enclosed mass increases. For this model, they obtain an SMBH mass of $M=2.249\times10^9  \Msun$. They report a statistical model-fitting uncertainty of $0.18\%$, systematic uncertainties of $0.62\%$, and $12\%$ from uncertainty in the distance to NGC 3258. Our black hole mass measurement is consistent with their favored model at the $0.25\sigma$ level.
    
    To best compare the molecular gas and stellar dynamical techniques, it makes sense to compare the models with as similar assumptions as possible.  To do this, we compare our results with their model D1, which uses an extended mass  profile based on their measured light profile with MGE deprojection, an assumed extinction correction of $A_H = 0$,  a parameterized $H$-band mass-to-light value, a flat disk structure, and a Gaussian turbulent velocity profile.  Although \citet{Boizelle2019} argue for significant extinction in the center, comparing this model to ours is the most straightforward, as they both make similar assumptions about the extended mass profile.  A key difference is that we assume a dark matter profile, which will change the effective mass-to-light ratio as a function of radius but predominantly at large radii.  This gives us a greater handle on the overall average stellar mass-to-light ratio, but the high-quality ALMA data allow \citet{Boizelle2019} to have a better handle on the extended mass distribution within $\sim1\arcsec$ of the galaxy's center.  
    
    Using model D1, \citet{Boizelle2019} find $M=2.276\times10^9  \Msun$ and $\Upsilon_{H} = 2.73 \MsunLsun$.  Our black hole masses are in strong agreement, though our mass-to-light ratios differ by $2.3\sigma$.  The spatial extent of their data, however, only encompasses the inner $\sim 300 \ \rm pc$ of NGC 3258. Our full data set covers approximately $3 \ \rm kpc$ and our models produce a mass-to-light ratio averaged over the full galaxy. Variations in the mass-to-light ratio across the radial profile are possible. To explore this, we examined the stellar template weights as a function of radius for our \CaII fits. We find that there is little to no variation in preferred stellar template as a function of radius. However, this method likely does not have the sensitivity to resolve variations of $\lesssim 15\%$ in the mass-to-light ratio.  
    
    Given that there is an obvious dust disk in this galaxy, assuming no extinction (as we do and as does model D1 in \citealt{Boizelle2019}) will result in different inferences about the mass-to-light ratio when the kinematic data sensitive to it come primarily from regions close to the disk (as do the ALMA data in \citealt{Boizelle2019}) or from the entire extent of the galaxy (as they do in our data).  It is expected that the ALMA data set would prefer a higher value of $\Upsilon_H$ than the MUSE data set as they are actually measures of different regions of the galaxy.  It is reassuring that despite the slight differences in assumptions, the black hole masses remain consistent.  \citet{Boizelle2019} note that in their preferred extinction model ($A_H = 0.75$) half of the stellar light within $R < 0\farcs25$ is absorbed, but this is a small fraction of the mass of the black hole.  The radius at which the mass in stars is equal to the black hole is at roughly 1\arcsec, and thus most of the stellar mass will be in a shell from approximately $0.75 < R < 1\arcsec$, where there is far less extinction.
    
    Previous works, such as \citet{Walsh2013}, have found discrepancies between gas and stellar dynamic mass (measured by \citealt{Gebhardt2011}) measurements; in this case, stellar dynamics produces a mass for the SMBH in M87 that is a factor of 2 greater than that measured by gas dynamics.  \citet{Osorno2023}, however,  found that the discrepancy, at least in part, can be explained by complexities in the morphology and kinematics of the nuclear ionized gas that were not resolved in the \textit{HST} spectroscopy used by \citet{Walsh2013}. These results, in combination with the results from this work, provide strong support for both the gas dynamical and stellar dynamical mass measurement methods.
    
    Since direct mass measurement methods are not always feasible, it is also important to compare our results to results obtained by indirect methods. We compare to the $M$--$\sigma$ and $M$--$L$ relations from \citet{Kormendy&Ho2013}. To compute the mass with the $M$--$\sigma$, we first calculate the effective velocity dispersion ($\sigma_e$) using the following equation:
\begin{equation} \label{eq: sig_e}
    \sigma_e^2 = \frac{\int_0^{R_e}(\sigma^2(r) + v^2(r))I(r)dr}{\int_0^{R_e}I(r)dr},
\end{equation}
    where $R_e$ is the half-light radius ($\approx66\arcsec$ or $10.6 \ \rm kpc$, adopted from \citealt{Boizelle2019}), $\sigma(r)$ is the velocity dispersion along the kinematic major axis, $v(r)$ is the velocity along the kinematic major axis, and $I(r)$ is the stellar surface brightness profile \citep{Tremaine2002,Gultekin2009}. Using our velocity dispersion values ($\sigma$) and their uncertainties computed in Section~\ref{sec: kin_extract}, we find an effective velocity dispersion of $\sigma_e = 232 \pm 7 \kms$, where the uncertainty is the $1\sigma$ confidence interval estimated from a Monte Carlo method. The $M$--$\sigma$ relation, defined as
\begin{equation} \label{eq: M-sigma}
    \frac{M}{10^9 \mathrm{M}_\odot} = \left(0.309^{+0.037}_{-0.033}\right)\left(\frac{\sigma_e}{200\,\mathrm{km\,s^{-1}}}\right)^{4.38 \pm 0.29},
\end{equation}
    predicts a mass of $\log(M/\rm M_{\scriptscriptstyle\odot}) = 8.8 \pm 0.3$, notably smaller than our best-fit $\log(M/\mathrm{M_{\scriptscriptstyle\odot}}) = 9.3$.

    Since we are adopting the surface brightness profile determined by \citet{Boizelle2019}, we adopt the same $K$-band absolute magnitude of $M_K = -24.33 \pm 0.45\ \rm mag$ \citep{Makarov2014}. Converting to $K$-band luminosity and using the \citet{Kormendy&Ho2013} $M$--$L$ relation, 
\begin{equation} \label{eq: M-L}
    \frac{M}{10^9 \mathrm{M}_\odot} = \left(0.542^{+0.069}_{-0.061}\right)\left(\frac{L_{K,\mathrm{bulge}}}{10^{11} \, \mathrm{L_{K\scriptscriptstyle\odot}}}\right)^{1.21 \pm 0.09},
\end{equation}
  predicts a mass of $M = (1.00^{+0.18}_{-0.16}) \times 10^9 \Msun$. Similar to the findings of \citet{Boizelle2019}, we find a significant discrepancy of $0.57 \, \rm dex$ between the SMBH mass estimates derived from the \citet{Kormendy&Ho2013} $M$--$\sigma$ and $M$--$L$ relations and the mass determined through stellar dynamics. The latter suggests a mass that is roughly $2$--$4$ times the estimated value. This difference exceeds the intrinsic scatter of these relations by about $0.27 \ \rm dex$. 
  
  For their $M$--$\sigma$ estimation, \citet{Boizelle2019} adopt a central velocity dispersion from the HyperLeda Database \citep{Makarov2014} of $260\pm10\kms$, which is an average of measurements taken from \citet{Davies1987}, \citet{Faber1989}, and \citet{Pellegrini1997}. These measurements were taken with optical observations with the Lick $3 \ \rm m$, Las Campanas Observatory $2.4 \ \rm m$, the Anglo-Austrian Telescope $3.9 \ \rm m$, and KPNO $2.1 \ \rm m$ telescopes and the Boiler \& Chivens spectrograph on the Cassegrain focus of the ESO $1.52 \ \rm m$ telescope. The authors quote instrumental dispersions ranging from $50$ to $215\kms$. The MUSE NFM has an instrumental dispersion of approximately $63\kms$ \citep{Simon2024}. Our measured central velocity dispersion, $\sigma_c = 302\pm12 \kms$ is approximately $16\%$ higher than the average of these measurements. Since the effective velocity dispersion $\sigma_e$ contains information about the stellar rotational velocities, we take this as our velocity dispersion measurement for use in the $M$--$\sigma$ relation.
    
\subsection{Consequences of the Cored Galaxy Profile}

    The most massive early-type galaxies typically display low-density cores \citep{Lauer2007}, likely formed from core scouring by binary SMBHs \citep{Ebisuzaki1991, Faber1997, Harris2024, Milosavljevic2001, Ravindranath2002, Thomas2014}. This low density results in a shallow, flat surface brightness profile near the inner part of the galaxy. Given the effective velocity dispersion ($\sigma_e$) dependence on the surface brightness profile, the core profile will suppress $\sigma_e$ and result in an underpredicted mass for the central SMBH. To explore the effects of core scouring in the surface brightness profile, we considered simple, plausible corrections to the surface brightness profile to add back in stellar content to approximate the surface brightness profile before a core-scouring event (i.e., we parameterize the surface brightness profile as a logarithmic profile). Computing the effective velocity dispersion with the updated surface brightness profile, we find a negligible suppression of only a few $\kms$. Therefore, the discrepancy in the $M$--$\sigma$ relation cannot be explained by the cored galaxy surface brightness profile alone. It should be noted, however, that we made no alterations to our measured $\sigma(r)$ values when recomputing $\sigma_e$. 
    
    In order to recover the mass measured with stellar dynamics, an effective velocity dispersion of approximately $310 \kms$ is required. While our central velocity dispersion is consistent with this value (and the central velocity dispersion and effective velocity dispersion are typically consistent with each other \citep{Kormendy&Ho2013}), we measure a significant difference in $\sigma_e$ and $\sigma_c$. This discrepancy can be attributed to two factors: 1) the mass of the SMBH in NGC 3258 is large, which inflates $\sigma_c$ when compared to $\sigma_e$, since $\sigma_e$ is more heavily weighted by the outer regions of the galaxy; and 2) NGC 3258 is a relatively nearby galaxy, which allows us to resolve stellar motions very close to the central SMBH, resulting in a higher central velocity dispersion as compared to more distant systems. 

\section{Summary} \label{sec: summary}

    In this work, we presented an analysis of the elliptical (E1) galaxy NGC 3258 to measure the mass of its central SMBH using stellar dynamics. Our data were acquired MUSE and combine IFU observations taken in both the WFM and NFM. We extracted the stellar kinematics by fitting the \CaII triplet in the NFM the \Mgb triplet in the WFM. Using these measurements, we construct LOSVDs, which are inputs for our models.

    To model the galaxy, we employed the three-integral axisymmetric Schwarzschild orbit library modeling method to fit the observed LOSVDs. From these models we obtained estimates for the SMBH mass ($M$), $H$-band mass-to-light ratio ($\Upsilon$), asymptotic circular velocity ($v_c$), and dark matter halo scale radius ($r_c$). Our derived mass of the SMBH is $(2.2 \pm 0.2)\times10^9 \Msun$. This value is consistent with a previous gas dynamical measurement by \citet{Boizelle2019} who found a mass of $2.249\times10^9 \Msun$ with 0.18\% statistical uncertainties 0.62\% systematic uncertainties. The close agreement between these results provides strong support for both the stellar dynamical and gas dynamical methods and serves as a reference point for future direct and indirect mass measurements of the SMBH in NGC 3258.
    
    We also compared our measurement to the mass predicted by SMBH--galaxy scaling relations. Using the $M$--$\sigma$ and $M$--$L$ relations, we find that our SMBH mass measurement is overmassive when compared to predicted values by $0.57 \ \rm dex$, which is greater than the intrinsic scatter of these relations ($\sim0.3 \ \rm dex)$. This discrepancy underscores the need for a better understanding of the systematic uncertainties governing the relations and how these relations evolve.
    
    This work contributes to the ongoing efforts to measure the masses of SMBHs and expand SMBH demographics, as well as to understand their coevolution with host galaxies and the relations between SMBH masses their global properties. Our findings serve as a benchmark for future studies using indirect methods and help address outstanding questions regarding SMBH demographics and their scaling relations.

\section*{Acknowledgments}

    T.K.W. thanks Naiara Patiño for the invaluable feedback on the text in this manuscript.
    N.N. and V.A. acknowledge funding from ANID Chile via Nucleo Milenio TITANs (NCN2023\textunderscore 002), Fondecyt 1221421, and BASAL FB210003.

    Based on observations collected at the European Southern Observatory under ESO program 105.20K2.

\software{Astropy \citep{astropy2013,astropy2018,astropy2022},
        Modelization of the Adaptive Optics Psf in PYthon \citep[MAOPPY;][]{Fetick2019,Beltramo2020},
        SAOImageDS9 \citep{Joye2003_ds9}, 
        SciPy\_python \citep{Oliphant2007}.}

\bibliography{references.bib}{}
\bibliographystyle{aasjournal}

\end{document}